\documentstyle[epsfig,floats,aps,preprint]{revtex}
\begin{document}
\draft
\preprint{\vbox{ \hbox{SOGANG-HEP 234/98} \hbox{SNUTP 98-020}
                 \hbox{hep-th/9803081} }}
\title{Initial singularity free quantum cosmology in two-dimensional \\
        Brans-Dicke theory}
\author{Won Tae Kim\footnote{electronic address:wtkim@ccs.sogang.ac.kr}
  and Myung Seok Yoon\footnote{electronic address:younms@physics.sogang.ac.kr}}
\address{Department of Physics and Basic Science Research Institute, \\ 
        Sogang University, Seoul
 121-742, Korea }
\date{March 1998}
\maketitle
\bigskip
\begin{abstract}
We consider two-dimensional Brans-Dicke theory to study the initial
singularity problem. It turns out that the initial curvature
singularity can be finite for a certain Brans-Dicke constant
$\omega$ by considering the quantum back reaction of the
geometry. For $\omega=1$, the universe starts with the finite
curvature scalar and evolves into the flat spacetime. Furthermore
the divergent gravitational coupling at the initial time 
can be finite effectively
with the help of quantum correction.  The
other type of universe is studied for the case of $0<\omega<1$.
\end{abstract}

\bigskip

\newpage

\noindent {\bf 1.~Introduction}\label{sec:intro}\\
The classical solution of general relativity
may yields curvature singularity in the black hole physics and cosmology.
This divergent physical quantity seems to be mild with the help of
quantum gravity. However, the consistent quantum theory
of gravity has not been established. The perturbation theory of
Einstein gravity is invalid in higher loops. As far as a consistent
and renormalizable quantum gravity does not appear, it seems to be
difficult to resolve the singularity problem.   
  
On the other hand, in two dimensions there exists some renormalizable
gravity such as Callan-Giddings--Harvey-Strominger (CGHS)
model~\cite{cghs} and Russo-Susskind-Thorlacius (RST) model~\cite{rst}
including soluble one-loop quantum effect. And various aspects of
quantum cosmology in two-dimensional gravity have been studied
in Ref.~\cite{mr,gv,nt,odi}. The two-dimensional Brans-Dicke theory
is also a good candidate to study the initial singularity problem
on the basis of conventional quantum field theory without encountering
the four-dimensional complexity. If the model gives the initial singularity
classically, it would be interesting to study how to modify the classical
singularity through the quantum back reaction of the geometry. 
If quantum corrections drastically modify the classical theory
and the curvature singularity does not appear, then the initial
curvature singularity may be ascribed to the classical concept. 

In this paper, we shall study the classical curvature singularity of
expanding universe in 
the two-dimensional Brans-Dicke model which exhibits the initial 
singularity of curvature scalar and the divergent coupling in sect. 2. 
This initial singularity
can be shown to be finite for some special value of Brans-Dicke constant
by considering the quantum back reaction of the geometry in sect. 3.
For $\omega =1$ the model exhibits the finite curvature scalar
at the initial comoving time whereas for $0<\omega <1$ the curvature
is bounded and the extremum exists at the finite comoving time.
And the gravitational coupling in both cases becomes finite.
If $\omega >1$, the curvature scalar is not bounded.
Finally discussion is given in sect.4.
\\
{\bf 2.~Classical Brans-Dicke theory}\label{sec:cl-BD}\\
We start with two-dimensional Brans-Dicke action, 
\begin{equation}\label{bd:act}
S_{\rm BD} = \frac{1}{2 \pi} \int\/d^{2}x\sqrt{-g}\, e^{-2\phi}
        \left[ R - 4\omega(\nabla\phi)^2 \right],
\end{equation}
where $R$ and $\phi$ are curvature scalar and redefined 
Brans-Dicke scalar field,
respectively, and $\omega$ is an arbitrary positive
constant. To write down $\psi= e^{-2\phi}$, then the
conventional Brans-Dicke form is recovered.
In our model, the large $\omega$
limit does not give the locally non-trivial gravity 
on the contrary to the four-dimensional Brans-Dicke theory 
since in two dimensions the Einstein-Hilbert action is proportional
to the Euler characteristic.

We now consider the classical
conformal matter fields given by
\begin{equation}\label{cl:act}
S_{\rm Cl} = - \frac{1}{2 \pi} \int\/d^{2}x\sqrt{-g} \,
        \frac12 \sum_{i=1}^N (\nabla f_i)^2,
\end{equation}
where $i=1,2,\dots,N$ and $N$ is the number of  conformal matter fields.
Then the
action~(\ref{bd:act}) and (\ref{cl:act})
lead to classical equations of motion with respect to $g_{\mu\nu}$ and $\phi$, 
\begin{eqnarray}
G_{\mu\nu} &=& T_{\mu\nu}^{\rm Cl}, \label{eq:cl-1} \\
R + 4\omega(\nabla\phi)^2 &-& 4 \omega \Box\phi = 0, \label{eq:phi}
\end{eqnarray}
where 
\begin{eqnarray}
G_{\mu\nu} &=& \frac{2\pi}{\sqrt{-g}} \frac{\delta S_{\rm BD}}{\delta
        g^{\mu\nu}} \nonumber \\
        &=& 2e^{-2\phi}\left\{\nabla_\mu\nabla_\nu\phi -
        2(1+\omega)\nabla_\mu\phi\nabla_\nu\phi + g_{\mu\nu}\left[
        (2+\omega)(\nabla\phi)^2 - \Box\phi\right]\right\} \label{eq:G}
\end{eqnarray}
and classical energy-momentum tensor $T_{\mu\nu}^{\rm Cl}$ sets to zero
for simplicity. 

We now
choose the conformal gauge such as $g_{\pm\mp}=-\frac12e^{2\rho}$ and
$g_{\pm\pm}=0$. From Eq.(\ref{eq:G}), we obtain the conformal gauge
fixed forms,
\begin{eqnarray}
G_{\pm\pm} &=& 2 e^{-2\phi}\left[ \partial_\pm^2\phi -
        2(1+\omega)(\partial_\pm\phi)^2 -
        2\partial_\pm\rho\partial_\pm\phi \right], \label{G:++} \\
G_{+-} &=& 2e^{-2\phi}\left[ 2\partial_+\phi\partial_-\phi -
        \partial_+\partial_-\phi \right], \label{G:+-}
\end{eqnarray}
and the equation of motion for $\phi$ is given by
\begin{equation} 
\partial_+\partial_-\rho - 2\omega\partial_+\phi\partial_-\phi +
2\omega\partial_+\partial_-\phi = 0,
\end{equation}
where $x^{\pm}=t \pm x$.
The classical solutions in the homogeneous space are given by 
\begin{eqnarray}
\rho &=& -\omega \phi,\label{rho} \\
e^{-2\phi} &=& Mt,
\end{eqnarray}
where $M$ is an integration constant and we take $M$ as a positive
constant to obtain positive time $t$. The other two integration
constants are chosen to be zero since they have 
no important role in our case.
On the other hand, in the comoving coordinates defined
by $ds^2 = - d\tau^2 + a^2(\tau)dx^2$,
the curvature scalar $R(\tau)$ and scale factor
$a(\tau)$ are written as follows, 
\begin{eqnarray}
R(\tau) &=& -\frac{4\omega}{(2+\omega)^2} \frac{1}{\tau^2}, \label{cl:R}\\
a(\tau) &=& \left[\left(1+\frac{\omega}{2}\right)
        M\tau\right]^{\frac{\omega}{2+\omega}}, \label{cl:a}
\end{eqnarray}
where $M\tau = \frac{2}{2+\omega} \left(Mt\right)^{1+\frac{\omega}{2}}$
($\tau>0$ and $t>0$).
It is clear that the curvature scalar has an initial singularity
with a power-law inflation. 
Since the behaviors of scale factor are 
$\dot{a}(\tau) \sim \tau^{-\frac{2}{2+\omega}} > 0$
and $\ddot{a}(\tau) \sim - \tau^{-\frac{4+\omega}{2+\omega}} < 0$,
and the universe shows the decelerating expansion. It is natural to
obtain the negative curvature scalar in Eq. (\ref{cl:R}) corresponding to 
the desirable decelerating universe 
since in two dimensions $R$ is directly proportional to $\ddot{a}({\tau})/a(\tau)$. 

Note that the gravitational coupling can be defined by 
\begin{equation}
g_{N}^2=e^{2\phi}
\end{equation}
and from Eqs. (\ref{rho}) and (\ref{cl:a}), it yields
\begin{equation}\label{eq:coupling1}
g_{N}^2= \left[\left(1+\frac{\omega}{2}\right)
        M\tau\right]^{-\frac{2}{2+\omega}}.
\end{equation} 
It shows that the divergent coupling at the initial time decreases and goes to
zero in the asymptotically flat spacetime.

In the next section, we study whether the initial curvature singularity
and the divergent coupling can be modified or not in the quantized theory. 
We hope the resulting
quantum theory gives the singularity free cosmology with the finite coupling.  
\\
{\bf 3.~Quantum Back Reaction}\label{sec:quantum}\\
We consider the one-loop effective action from the conformal
matter fields which is given by
\begin{equation}
S_{\rm Qt} = \frac{\kappa}{2\pi}\int d^2x\/ \sqrt{-g}\left[
        -\frac14 R\frac{1}{\Box}R - \omega\left(\nabla\phi\right)^2
        - \frac{1-\omega}{2}\phi R\right],\label{qt:act-origin}
\end{equation}
where $\kappa=\hbar\frac{N-24}{12}$. The first term in
Eq. (\ref{qt:act-origin}) is due to the induced gravity from the
conformal matter fields. The
other two terms are regarded as local regularization ambiguities of
conformal anomaly. The similar effective action was already treated on
the purpose of studying the graceful exit problem of the string
cosmology in the large $N$ limit~\cite{bk,wy:ten}.
The higher order of quantum correction beyond one loop is neglizable
in the large $N$ approximation where $N \rightarrow \infty$ and 
$\hbar \rightarrow 0$ so that $\kappa$ is assumed to be positive finite
quantity.

By introducing an auxiliary field $\psi$, 
Eq.(\ref{qt:act-origin}) may be written in the local form of
\begin{equation}
S_{\rm Qt} = \frac{\kappa}{2\pi}\int d^2 x \/ \sqrt{-g}\left[
        \frac14 R\psi-\frac{1}{16}(\nabla\psi)^2
        -\omega\left(\nabla\phi\right)^2
        - \frac{1-\omega}{2}\phi R\right].\label{qt:nonlocal}
\end{equation}
Now the total effective action is defined by 
\begin{equation}\label{eq:eff.act}
S_{\rm T} = S_{\rm BD} + S_{\rm M}, \label{total}
\end{equation}
where the matter part is formally composed of two pieces of $S_{\rm M}=S_{\rm
Cl} + S_{\rm Qt}$, however, we neglect the conformal matter fields just like
the classical cosmology for simplicity. The equations of motion with
respect to $g_{\mu\nu}$, $\phi$, and $\psi$ from the action (\ref{eq:eff.act}) are
\begin{eqnarray}
G_{\mu\nu} &=& T_{\mu\nu}^{\rm M}, \label{G=T} \\
e^{-2\phi}\left[R-4\omega\Box\phi+4\omega(\nabla\phi)^2 \right]
         &=& -\frac{\kappa}{4}(1-\omega)R+\omega\kappa\Box\phi, \label{cov:phi}\\
\Box \psi &=& -2R, \label{cov:psi}
\end{eqnarray}
where $T_{\mu\nu}^{\rm M}$ is the energy-momentum tensor, which is
given by
\begin{eqnarray}
& &T_{\mu\nu}^{\rm M} \ = -\frac{2\pi}{\sqrt{-g}}
        \frac{\delta S_{\rm M}}{\delta g^{\mu\nu}} \nonumber \\
& & \qquad \, =
        \frac{\kappa}{4}\left[\nabla_\mu\nabla_\nu\psi
        + \frac14\nabla_\mu\psi\nabla_\nu\psi
        - g_{\mu\nu}\left(\Box\psi+\frac18(\nabla\psi)^2
        \right)\right] \nonumber \\
& & \qquad \ \ \ -\frac{\kappa}{2}(1-\omega)\bigg[
        \nabla_\mu\nabla_\nu\phi - g_{\mu\nu}\Box\phi\bigg]
        + \kappa\omega\left[\nabla_\mu\phi\nabla_\nu\phi
        - \frac12 g_{\mu\nu}(\nabla\phi)^2\right].
        \label{cov:T} 
\end{eqnarray}

In the conformal gauge fixing, the energy-momentum tensor is written as
\begin{eqnarray}
T_{\pm\pm}^{\rm M} &=& \kappa\left[ \partial_\pm^2\rho -
        (\partial_\pm\rho)^2 \right] - \frac{\kappa}{2}(1-\omega)\left[
        \partial_\pm^2\phi - 2\partial_\pm\rho\partial_\pm\phi \right]
        + \kappa\omega(\partial_\pm\phi)^2 - \kappa t_\pm, \label{qt:T++}\\
T_{+-}^{\rm M} &=& -\kappa\partial_+\partial_-\rho +
        \frac{\kappa}{2}(1-\omega)\partial_+\partial_-\phi, \label{qt:T+-}
\end{eqnarray}
where $t_{\pm}$ arises from elimination of auxiliary field and 
reflects the nonlocality of induced gravity.
Defining new fields as
follows~\cite{wy:ten,wl,bc},
\begin{eqnarray}
\Omega &=& -\frac{\kappa}{2}(1 + \omega)\phi
        + e^{-2\phi}, \label{Omega} \\
\chi &=& \kappa\rho - \frac{\kappa}{2}(1-\omega)\phi
        + e^{-2\phi},\label{chi}
\end{eqnarray}
the equations of motion (\ref{G=T})--(\ref{cov:psi}) are obtained in the simple form of
\begin{eqnarray}
\partial_+\partial_-\Omega &=& 0, \label{eq:Omega}\\
\partial_+\partial_-\chi &=& 0, \label{eq:chi}
\end{eqnarray}
and the constraints are given by
\begin{eqnarray}
G_{\pm\pm} -T_{\pm\pm}^{\rm M} &=& -\frac{1}{\kappa}\left(
        \partial_\pm\Omega\right)^2 + \frac{1}{\kappa}
        \left(\partial_\pm\chi\right)^2 - \partial_\pm^2\chi
        + \kappa t_\pm \nonumber \\
        &=& 0. \label{qt:constr.}
\end{eqnarray}
In the homogeneous spacetime, general solutions are
\begin{eqnarray}
\chi &=& \chi_0 t + A, \label{sol:chi}\\
\Omega &=& \Omega_0 t + B, \label{sol:Omega}
\end{eqnarray}
where $\Omega_0$, $\chi_0$, $A$, and $B$ are constants. Choosing the
quantum matter state as vacuum~\cite{rey}, $t_\pm = 0$, the constraint
equation (\ref{qt:constr.}) results in $\Omega_0 = \pm
\chi_0$. Hereafter we consider only the case of $\Omega_0 = + \chi_0
\equiv M$ corresponding to the classical solution
($\rho=-\omega\phi$) for $\kappa \rightarrow 0$. Also we can take $A=B=0$ without loss of
physical result. From the definitions of $\chi$ and $\Omega$
in Eqs.~(\ref{Omega}) and (\ref{chi}) and the general solutions
of Eqs.~(\ref{sol:chi}) and (\ref{sol:Omega}), we obtain the following
closed forms
\begin{eqnarray}
& & e^{\frac{2}{\omega}\rho} + \frac{1}{2\omega}( 1 +
        \omega)\kappa\rho = M t, \label{sol:rho} \\
& & e^{-2\phi} - \frac12(1 + \omega )\kappa\phi = Mt.
        \label{sol:phi} 
\end{eqnarray}

To study how the universe evolves as time goes on, we redefine
time $t$ as a comoving time $\tau$ defined by $\tau = \int^t dt \/
e^{\rho(t)}$, and then the metric can be expressed as $ds^2 = -
e^{\rho(t)} ( dt^2 - dx^2 ) = - d\tau^2 + a^2 (\tau) dx^2$, where
$a(\tau)$ is a scale factor.
In the case of $\tau \rightarrow +0$ ($t \rightarrow
-\infty$), the behavior of scale factor is approximately
given by
\begin{eqnarray}
& &a(\tau) \approx \frac{2\omega M}{\kappa(1 + \omega)} \tau
        \rightarrow +0, \label{a:-} \\
& &\dot{a}(\tau) \approx \frac{2\omega M}{\kappa(1 + \omega)}
        > 0, \label{da:-} \\
& &\ddot{a}(\tau) \approx - \frac{\omega}{(1+\omega)^3}
        \frac{32}{\kappa^3} M^2
        \left( \frac{2\omega M}{\kappa(1 + \omega)} \tau
        \right)^{\frac{2-\omega}{\omega}} < 0, \label{dda:-} 
\end{eqnarray}
where $\tau \approx \frac{(1+\omega)\kappa}{2\omega M}
e^{\frac{2\omega M}{(1+\omega)\kappa} t}$. 
Hence in initial stage of
inflation, the size of the universe approaches zero
and the universe exhibits the decelerating expansion. 
And the asymptotic behavior of scale factor
for  $\tau \rightarrow +\infty$
(i.e. $t \rightarrow +\infty$) is given by
\begin{eqnarray}
& &a(\tau) \approx \left[ \left( 1 + \frac{\omega}{2} \right) M \tau
        \right]^{\frac{\omega}{2+\omega}} \rightarrow +\infty,
        \label{a:+} \\
& &\dot{a}(\tau) \approx \frac{\omega}{2}M \left\{ \left[ \left(1 +
        \frac{\omega}{2} \right) M \tau \right]^{\frac{2}{2+\omega}} +
        \frac{\kappa}{4}(1+\omega) \right\}^{-1} \rightarrow + 0,
        \label{da:+} \\
& &\ddot{a}(\tau) \approx -\frac{\omega}{2}M^2 \left[ \left( 1 +
        \frac{\omega}{2} \right)M\tau
        \right]^{-\frac{4+\omega}{2+\omega}} \rightarrow -0,
        \label{dda:+} 
\end{eqnarray}
where $\tau \approx \frac{2}{M(2+\omega)} (Mt)^{\frac{2+\omega}{2}}$.
This shows that the spacetime exhibits the decelerating expansion for
$0<\tau<\infty$.
\begin{figure}[h,t]
\vspace{1.3cm}
\centerline{\hbox{
    \psfig{figure=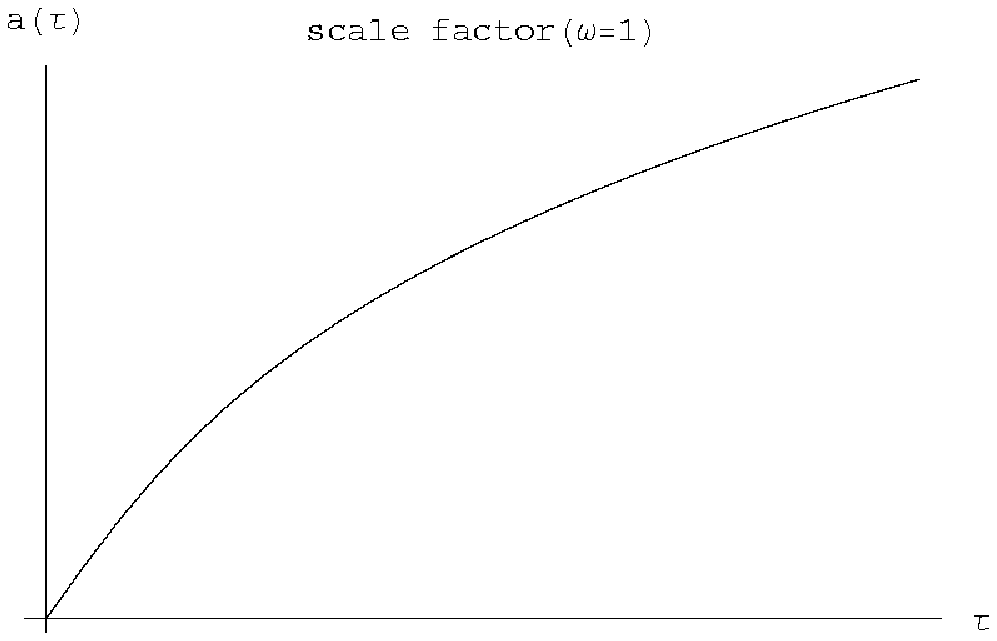,width=7cm,height=5cm}\hspace{1cm}
    \psfig{figure=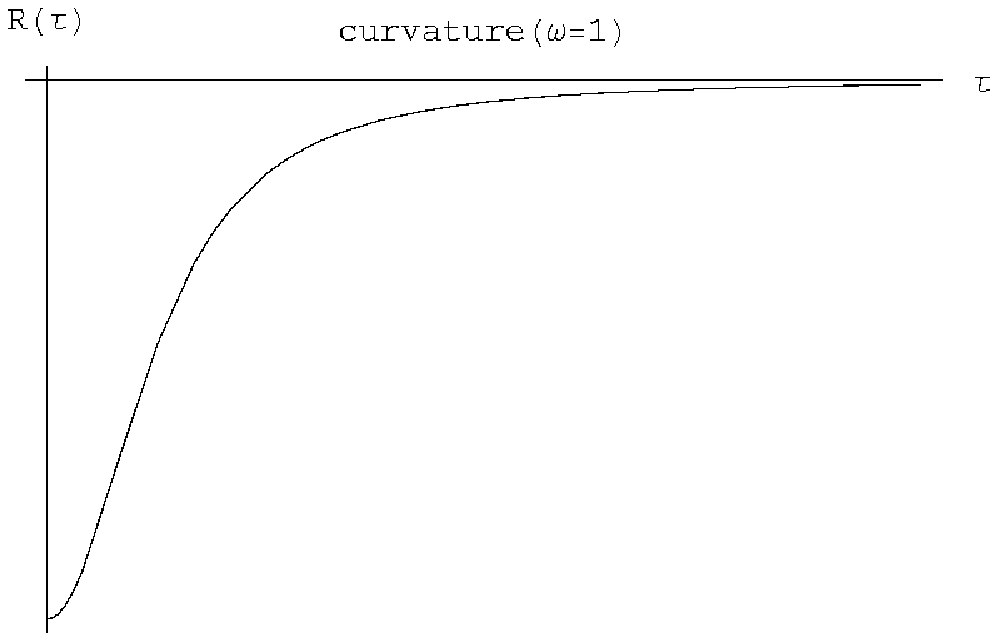,width=7cm,height=5cm}}}
\vspace{0.2cm}
\caption{For $\omega=1$, the curvature scalar in the beginning of inflation
approaches a constant value and spacetime is flat in the
far future.}
\label{omega=1}
\end{figure}
\begin{figure}[h,t]
\vspace{1.3cm}
\centerline{\hbox{
    \psfig{figure=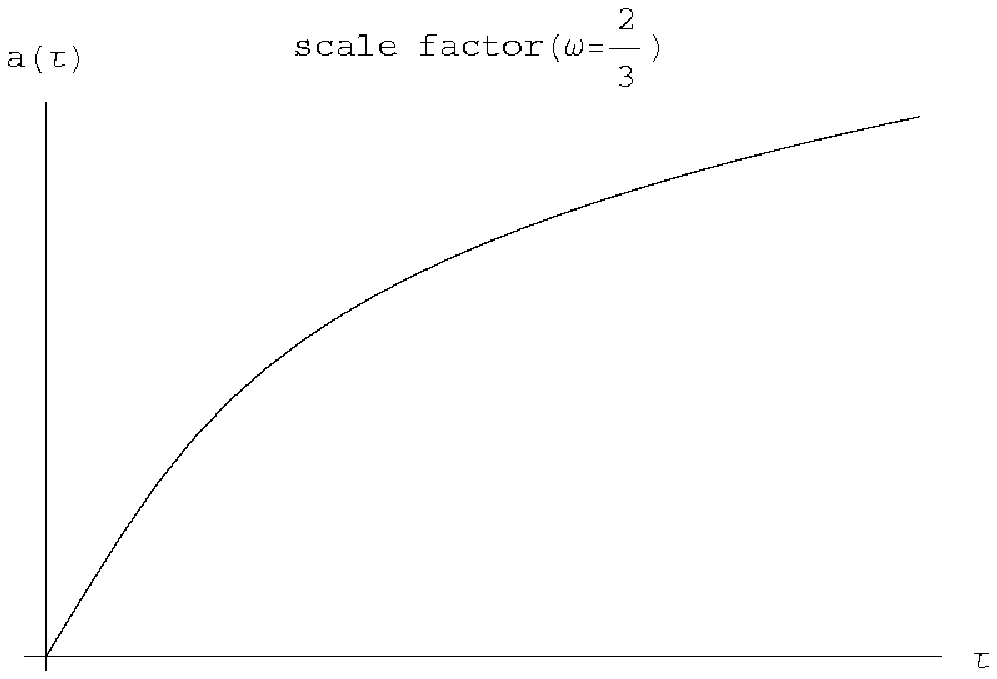,width=7cm,height=5cm}\hspace{1cm}
    \psfig{figure=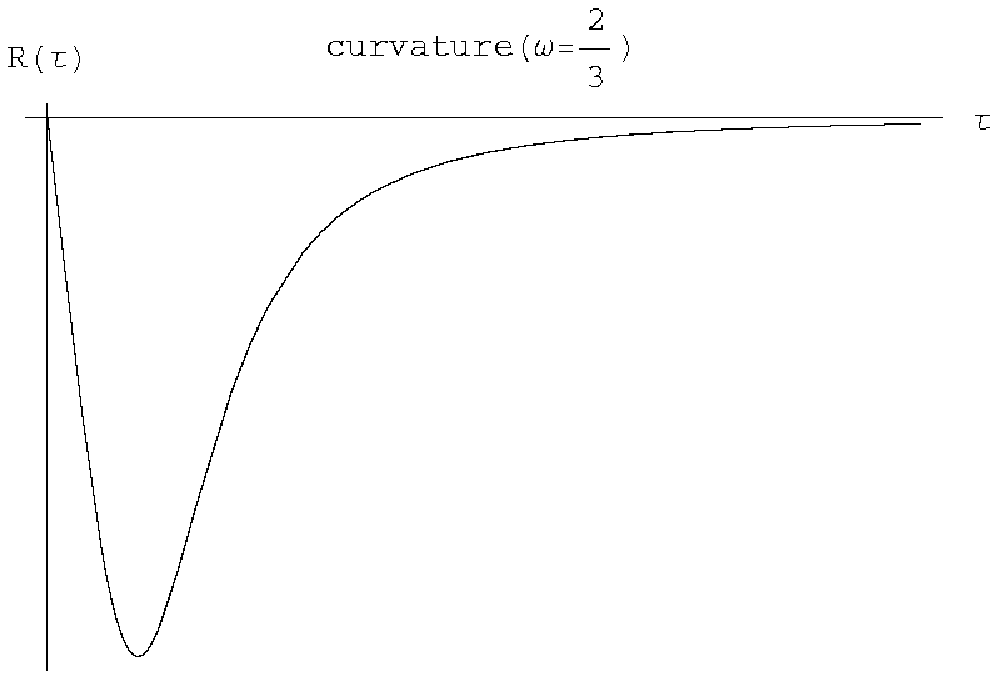,width=7cm,height=5cm}}}
\vspace{0.2cm}
\caption{For $\omega=\frac23$, the curvature scalar in the beginning
of inflation approaches zero and spacetime is flat in
the far future.}
\label{omega=2/3}
\end{figure}

The exact expression of curvature
scalar is written in the form of
\begin{equation}\label{R:exact}
R(\tau) = -\omega M^2 a^{\frac{2}{\omega}(1-\omega)}(\tau) \left[
a^{\frac{2}{\omega}}(\tau) + \frac{\kappa}{4}(1+\omega) \right]^{-3}.
\end{equation}
Note that the condition for boundedness of the scalar curvature is
given by the range of $0<\omega \le 1$ 
as easily seen from Eq.(\ref{R:exact}).
Therefore we restrict the Brans-Dicke constant as $0<\omega \le 1$
to avoid curvature singularity. 
The initial curvature scalar approaches asymptotically 
zero for $0< \omega < 1$.
It is finite $R(0)=-\frac{8M^2}{\kappa^3}$ 
for $\omega=1$ (see FIG.~\ref{omega=1} and \ref{omega=2/3}).

In fact, we can obtain the closed form of scale factor. By differentiating
Eq.(\ref{sol:rho}) and transforming coordinates from $(t,x)$ to
$(\tau,x)$, we find the equation of scale factor $a(\tau)$:
\begin{equation}\label{mot:a}
\dot{a}({\tau}) = \frac{\omega}{2} M \left[
a^{\frac{2}{\omega}}(\tau) + \frac{\kappa}{4}(1 + \omega)
\right]^{-1}.
\end{equation}
Through the integration of Eq.(\ref{mot:a}), it leads to the
closed form
\begin{equation}\label{mot2:a}
\left[a(\tau)\right]^{1+\frac{2}{\omega}} + \frac{\kappa}{4}(1 +
\omega)\left(1 + \frac{2}{\omega}\right)a(\tau) = \left(1 +
\frac{\omega}{2}\right) M \tau 
\end{equation}
with the initial condition $a(0)=0$. 
It is difficult to generically show the behavior of scale factor, 
so the scale factor and curvature scalar are depicted for 
the special values $\omega=1$ and $\omega=\frac{2}{3}$ in
FIG.~\ref{omega=1} and \ref{omega=2/3}.

As for the gravitational coupling, it is easily seen from the string theoretic
point of view by using the nonlinear sigma model in~\cite{rt}. 
The total action (\ref{total}) can be written in the form of
\begin{equation}
S_{\rm{T}}=\int d^2 \sigma G_{ij}(X) \partial X^i \partial X^j
\end{equation}
where the target metric is given by
\begin{equation}
  G_{ij} = \left( 
  \begin{array}{cc}
    2 \omega \left( e^{-2\phi} + \frac{\kappa}{4} \right) & 
    e^{-2\phi} + \frac{\kappa}{4}(1-\omega) \\
    e^{-2\phi} + \frac{\kappa}{4}(1-\omega) & -\frac{\kappa}{2} 
  \end{array}\right) \label{eq:Gij}
\end{equation}
and the target coordinate is $X^i=(\phi,\rho)$.
Therefore we obtain the effective coupling $g_{\rm eff}$ as
\begin{equation}
  \label{eq:g_eff}
  g_{\rm eff}^2 = \frac{g_N^2}{ 1+\frac{\kappa}{4}(1+\omega)g_N^2}.
\end{equation}
In the classical theory such as $\kappa=0$
($g_{\rm eff} = g_{N}$), the coupling 
diverges at the initial time of inflation as seen from
Eq.~(\ref{eq:coupling1}) and
approaches zero in the far future.
However, after taking into account of
the quantum back reaction, the coupling is interestingly given by the finite value
$\frac{4}{\kappa(1+\omega)}$ in the beginning of inflation and decreases
monotonically during the expansion of the universe.

Next we study what is the source of the decelerating expansion of
universe.
If we take the quantum mechanically induced matter as a 
perfect fluid, then we can realize the induced pressure is directly
related to the curvature scalar. And the induced energy is always
zero from the constraint equation. To show this fact, 
Eqs.~(\ref{eq:G}) and (\ref{cov:T}) 
are written in the comoving coordinate as
\begin{eqnarray}
G_{\tau\tau} &=& T_{\tau\tau}^{\rm M} = 0, \label{energy}\\
G_{xx} &=& - \frac{2}{\omega}[a(\tau)]^{2+\frac{2}{\omega}} \left[
        \frac{\ddot{a}(\tau)}{a(\tau)} + \frac{2}{\omega}\left(
        \frac{\dot{a}(\tau)}{a(\tau)}\right)^2 \right], \label{G:xx}\\
T_{xx}^{\rm M} &=& \frac{\kappa}{2\omega}
        (1+\omega)a(\tau)\ddot{a}(\tau) \label{T:xx}
\end{eqnarray}
by using the solution
$\phi(\tau) = -\frac{1}{\omega} \ln a(\tau)$.
Then we obtain
\begin{equation}
     T_{xx}^{\rm M} 
        = -\frac{\kappa}{4} (1+\omega)M^2 a^{\frac{2}{\omega}}(\tau)
        \left[ a^{\frac{2}{\omega}}(\tau) + \frac{\kappa}{4}(
        1+\omega) \right]^{-3} \label{T:xx-sol}
\end{equation}
after eliminating $\ddot{a}(\tau)$ by using  Eq.(\ref{mot:a}).
The pressure for the perfect fluid becomes
\begin{eqnarray}
p &\equiv& \frac{1}{a^2}T_{xx}^{\rm M} \\ \nonumber
  &=& \frac{\kappa}{4}(1+\frac{1}{\omega})R,
\end{eqnarray}
where we used Eq.~(\ref{R:exact}).
Note that the curvature scalar which characterizes the geometry
in two dimensions is of relevance to the pressure. This is a dynamical
equation of motion while Eq. (\ref{energy}) is just a constraint
equation in the comoving coordinate. Therefore, the source of dynamical 
evolution of the geometry is determined by the pressure. In our model,
the induced energy always vanishes.

\noindent{\bf 4.~Discussion}\label{sec:discuss}\\
The curvature scalar is defined by 
$R=\frac{2\ddot{a}(\tau)}{a(\tau)}$ in two-dimensional homogeneous space.
In our model, the scale factor $a(\tau)$ vanishes at 
$\tau \rightarrow +0$.
So one might be wonder why the curvature scalar is finite for $(\omega=1)$
or zero for $(0 <\omega \le 1)$ at $\tau \rightarrow +0$.
If the scale factor is expanded as 
$a(\tau)=a_1\tau +\frac12 a_2 \tau^2+ O({\tau}^3)$, then the scalar curvature diverges unless $a_2=0$.
So the finiteness of curvature requires the
absence of the order of ${\tau}^2$ in the asymptotic expansion of $a(\tau)$
around $\tau=0$. Let us exhibit two special cases of $\omega=1$ and
$\omega=\frac23$ for simplicity. Then the scale factors around the
initial comoving time are expanded as 
\begin{eqnarray}
a(\tau) &=& \frac{M}{\kappa}\tau - \frac{2M^3}{3\kappa^4}\tau^3 +
        O(\tau^4) \qquad \qquad \mbox{if $\omega=1$} \label{series:a1}, \\
a(\tau) &=& \frac{4M}{5\kappa}\tau - \frac{3(4M)^4}{(5\kappa)^5}\tau^4
        + O(\tau^5)\qquad \mbox{if $\omega=\frac23$} \label{series:a23}.
\end{eqnarray}
This is the reason how the curvature scalar is finite or zero near the
origin although the scale factor goes to zero.

In summary, we have studied the curvature singularity problem using
the conventional quantum field theory in the two-dimensional Brans-Dicke
cosmology and obtained the bounded curvature scalar and the finite gravitational
coupling
for $0<\omega \le 1$. 
We hope that the consistent quantum gravity may solve the singularity
problems in the realistic cases in the future.
\\
{\bf Acknowledgments}\\
W. T. Kim is very grateful to R. H. Brandenberger 
for useful discussions. 
This work was supported by Ministry of Education, 1997, Project
No. BSRI-97-2414, and Korea Science and Engineering Foundation through
the Center for Theoretical Physics in Seoul National University(1998).


\end{document}